\documentclass[aps,preprint]{revtex4}%
\usepackage{amsfonts}
\usepackage{amsmath}
\usepackage{amssymb}
\usepackage{graphicx}%
\providecommand{\U}[1]{\protect\rule{.1in}{.1in}}

\begin{document}
\title{Interface states in bilayer graphene and valleytronics}
\author{M. Ahsan Zeb}
\email{maz24@cam.ac.uk}
\affiliation{Department of Earth Sciences, University of Cambridge, UK.}

\begin{abstract}
We study the states localized near an interface between conducting and
insulating bilayer graphene (BLG) and show that they have highly unusual
properties that have no analog in conventional systems. Moreover, the states
belonging to the two independent valleys in the Brillouin zone of BLG show
contrasting properties that allow an \emph{easy} experimental realization of
various valley based functionalities desired in valleytronics \emph{without}
requiring any sophisticated techniques.

\end{abstract}
\date{22 Aug. 2010}
\maketitle

Bilayer graphene--- two layers of carbon atoms with the same intralayer
arrangement of atoms and relative stacking of the layers as in natural
graphite, is focus of intensive experimental and theoretical research these
days due to its unusual physical properties\cite{c1,c2,c3,c4,c5,c6,c7,c8}.
There are two independent valleys in Brillouin zone of BLG and the low energy
elementary excitations in both valleys posses a pseudospin and have
pseudospinor wavefunctions. The valley degree of freedom can be used for
controlling an electronic device similar to the spin or charge and there are
many proposals for achieving basic functionalities desired in this field,
named valleytronics, using graphene systems\cite{c9,c10,c11,c12,c13,c14}. In
ref\cite{c10}, A. Rycerz et. al. proposed a valley filter that is a strip of
monolayer graphene containing a narrow constriction with zigzag edges.
However, experimental realization of such a device is challenging, if not
impossible. Taking advantage of the effects of trigonal warping at high
carrier densities in monolayer graphene, J. L. Garcia-Pomar et. al.\cite{c11}
proposed a device which again requires zigzag edges in addition to
manipulation of beams of particles$-$ again a great challenge to the
experiment. In bilayer graphene, A. S. Moskalenko and J. Berakdar\cite{c12}
showed that intense \emph{shaped} light pulse can induce valley polarized
currents and, D. S. L. Abergel and Tapash Chakraborty\cite{c13} proposed a
device for generating valley polarized currents where in \ the presence of an
intense terahertz light source in a region of bilayer graphene with a finite
band gap the dynamically induced states can be made to exist only in a given
valley by tuning different parameters. Although, these are relatively easy for
experimental tests, they are hard to implement in practical devices. H.
Schomerus' proposal\cite{c14} uses the fact that the angular dependence of the
transmission probabilities of the particles of the two valleys from one region
to another differ when the two regions have opposite pseudospin polarization.
It demands a relatively sharp fermi surface and again controlling beams of
particles and their propagation angles. Here we show that states that are
localized near an interface of zero band gap and finite band gap BLG have
highly unusual properties that provide us with a conceptually extremely simple
way of obtaining valley polarization and related basic functionalities, and
eliminate almost all experimental challenges that stand in the way of
valleytronic devices.

In a clean zero band gap or conducting BLG at any energy there are two
pseudospinor plane wave states and two pseudospinor evanescent wave states.
The latter have appreciable magnitudes only near interfaces between regions of
different potentials and play an important role in transport properties.
Similar is the case of a clean finite band gap or insulating BLG where the
evanescent waves are localized near interfaces between regions of different
band gaps or potentials. For the first case they are only present at energies
which are outside the band gap of at least one side. For energies outside band
gap these states have the usual properties, however, for energies inside the
band gap, their wavevectors become complex so the spatial parts of their
wavefunctions contain plane waves as well as evanescent ones. This is unusual
for any wave in a pristine and dissipation-free system and here it is more
unusual as it apparently shows that number of particles is not conserved. In
search of resolution to this issue we derive the continuity equation
associated with the Hamiltonian of the system and check the current these
states carry along the decay direction. It comes out to be \emph{zero} just
like the usual evanescent waves which is quite interesting. However, these
states show a more interesting behavior in the transverse direction: the
states belonging to the two valleys carry currents in \emph{opposite}
and\emph{ fixed} directions,\emph{ independent} of the sign of the wavevector
component along this direction. The latter property is in stark contrast to
common systems described by the usual Schrodinger wave equation. For an
interface of a conducting and insulating BLG, due to the continuity of
probability current density, these states modifies the behavior of the states
localized on the conducting side as well. Thus the particles of a given valley
in conducting region at \emph{any energy} inside the band gap of the
insulating BLG coming towards the interface at \emph{any angle} turn towards
\emph{the same} lateral side close to the interface. Thus without worrying
about the propagation angles of the quasiparticles or the finite smearing of
the fermi surface due to temperature or any other intravalley scatterings, we
can easily separate particles of the two valleys and also perform other valley
based operations just by using a simple junction of conducting and insulating
BLG either in above mentioned geometry or by applying a voltage difference
along it as proposed in ref\cite{c9} for the topologically confined zero
energy chiral modes.

The low energy electronics properties of the BLG system lying in the xy-plane
can be approximately described by effective Hamiltonians $H^{\pm}$ for the two
valleys $K(+)$ and $K^{\prime}(-)$ \cite{c7,c8} which along with their
pseudospinor eigenfunctions $\Psi^{\pm}(x,y)$ can be written as%

\begin{equation}
H^{\pm}=%
\begin{pmatrix}
-U & \nabla_{\mp}\\
\nabla_{\pm} & U
\end{pmatrix}
,\Psi^{\pm}(x,y)=%
\begin{pmatrix}
\varphi_{A}^{\pm}\\
\varphi_{B}^{\pm}%
\end{pmatrix}
\label{H,psi}%
\end{equation}

Where the components $\varphi_{A}^{\pm}$ and $\varphi_{B}^{\pm}$ of $\Psi
^{\pm}(x,y)$ are envelope functions that give the amplitudes on the two
different layers on sites A and B that do not lie directly above or below
sites on the other layer. $\nabla_{\pm}=\frac{\hbar^{2}}{2m}(\partial_{x}\pm
i\partial_{y})^{2}$ where $\partial_{x,y}$ are differential operators, the two
layers have on-site energies equal to $\pm U$ induced by applied electrostatic
gates, and $m$ is the effective mass of quasiparticles when $U=0$. When
$U\neq0$ inversion symmetry between the two layers is broken and a band gap
equal to $2\left\vert U\right\vert $ is produced. Dispersion for both valleys
is the same and reads: $E(\mathbf{k})=\pm\sqrt{U^{2}+(\frac{\hbar^{2}k^{2}%
}{2m})^{2}}$ where $\pm$ refer to electron and hole bands here. Suppose an
interface between conducting region (CR, $x<0$) and insulating region (IR,
$x>0$) of BLG at $x=0$ then $U=0$ for $x<0$ and we have continuum of
propagating and evanescent states in CR while in IR for energies inside the
gap only the states localized near the interface are present. The states in CR
match those in the IR with the same energy $E$ and, due to the invariance
along the interface, with the same y-component of wavevector $k_{y}$ to ensure
the continuity of the probability density and the probability flux density.
Thus we can use $E$ and $k_{y}$ to label all the states. In IR, for a given
$E$ and $k_{y}$, the x-components of wavevectors $\pm k_{x}$ are complex
quantities given by $k_{x}=\pm k_{r}+ik_{i}$ where $k_{r}=\eta\sin\alpha$ and
$k_{i}=$ $\eta\cos\alpha$, and we defined $\eta=(\delta^{2}+k_{y}^{4})^{1/4}$,
$\alpha=\frac{1}{2}\tan^{-1}(\delta/k_{y}^{2})$ and $\delta=\sqrt{U^{2}-E^{2}%
}$. For simplicity we absorbed $\frac{2m}{\hbar^{2}}\ $in energy terms and we
will do so hereafter. This results in \emph{decaying} \emph{plane waves}$.$The
wavefunctions of the states belonging to the two valleys are $\Psi_{2}^{\pm
}(x,y)=\Psi_{2}^{\pm}(x)e^{ik_{y}y}$ where, ignoring the solutions that
diverge with $x$, $\Psi_{2}^{\pm}(x)$ are given by%

\begin{equation}
\Psi_{2}^{\pm}(x)=a_{2}^{\pm}%
\begin{pmatrix}
\frac{E-U}{\Omega_{\pm}}\\
-\frac{R_{\pm}}{\Omega_{\pm}}e^{2i\theta_{\pm}}%
\end{pmatrix}
e^{ik_{r}x-k_{i}x}+d_{2}^{\pm}%
\begin{pmatrix}
\frac{E-U}{\Omega_{\pm}}\\
-\frac{R_{\pm}}{\Omega_{\pm}}e^{-2i\theta_{\pm}}%
\end{pmatrix}
e^{-ik_{r}x-k_{i}x} \label{wf2}%
\end{equation}

Here $a_{2}^{\pm}$ and $d_{2}^{\pm}$ are complex coefficients determined by
matching conditions, $R_{\pm}=k_{r}{}^{2}+(k_{i}\pm k_{y})^{2}$, $\Omega_{\pm
}=\sqrt{(E-U)^{2}+R_{\pm}^{2}}$ and $\theta_{\pm}=\tan^{-1}(\frac{k_{i}\pm
k_{y}}{k_{r}})$. The spatial parts of the above wavefunctions apparently show
currents along $\pm x$ with magnitudes decreasing with $x$, however, as we
will shortly confirm, the relative phases $\pm2\theta_{\pm}$ between upper and
lower components of the pseudospinors contain the information needed to
produce the proper results. We derive expressions for probability current
densities $\mathbf{J}^{\pm}$ for the two valleys associated with the
Hamiltonians $H^{\pm}$ in equation\ref{H,psi}, they read: $\mathbf{J}^{\pm
}=(J_{x}^{\pm},J_{y}^{\pm})$ where $J_{x}^{\pm}$ and $J_{y}^{\pm}$ are given
by
\begin{align*}
J_{x}^{\pm}  &  =\frac{-\hbar}{m}\{\operatorname{Im}(\Psi^{\dagger}\sigma
_{x}\partial_{x}\Psi\pm\Psi^{\dagger}\sigma_{y}\partial_{y}\Psi)\}\\
J_{y}^{\pm}  &  =\frac{-\hbar}{m}\{\operatorname{Im}(-\Psi^{\dagger}\sigma
_{x}\partial_{y}\Psi\pm\Psi^{\dagger}\sigma_{y}\partial_{x}\Psi)\}
\end{align*}

Here $\sigma_{x,y}$ are Pauli's matrices. It is also worth mentioning that
these expressions show that unlike common systems the continuity of flux
density is ensured with that of wavefunction and its first derivative in spite
of the fact that in case of finite band gap BLG we have energy dependant
effective mass as is clear from the dispersion relation. Further, the mixed
$x,y$ terms shows some intimate link between the two orthogonal directions.
Using these expressions and defining $\beta=\tan^{-1}(\frac{k_{y}}{k_{r}})$
the angle that $\mathbf{q=(}k_{r},k_{y}\mathbf{)}$ makes with the x-axis, it
is straightforward to show that the currents $\mathbf{J}_{2}^{\pm}%
=(J_{2x}^{\pm},J_{2y}^{\pm})$ carried by $\Psi_{2}^{\pm}(x)$ are
\begin{align*}
J_{2x}^{\pm}  &  =-C^{\pm}(\left\vert a_{2}^{\pm}\right\vert ^{2}-\left\vert
d_{2}^{\pm}\right\vert ^{2})q\cos(2\theta_{\pm}\mp\beta)e^{-2k_{i}x}\\
J_{2y}^{\pm}  &  =C^{\pm}\{\mp(\left\vert a_{2}^{\pm}\right\vert
^{2}+\left\vert d_{2}^{\pm}\right\vert ^{2})q\sin(2\theta_{\pm}\mp
\beta)+2k_{y}\operatorname{Re}(a_{2}^{\pm}d_{2}^{\pm\ast}e^{2i(\theta_{\pm
}+k_{r}x)})\ \}e^{-2k_{i}x}%
\end{align*}

where $C^{\pm}=\frac{\hbar}{m}\frac{(U-E)R_{\pm}}{\Omega_{\pm}^{2}}$,
$q=\left\vert \mathbf{q}\right\vert $ and * shows complex conjugation. Note
that the cross terms in $J_{2x}^{\pm}$ vanish. \bigskip Using a few common
trigonometric identities we get $2\theta_{\pm}\mp\beta=\pi/2$ independent of
any parameter so \emph{both} terms in $J_{x}^{\pm a}\ $vanish to give
\[
J_{x}^{\pm a}=0
\]
Thus the relative phases between components of pseudospinors count for the
presence of plane waves and we obtain sensible results. A more interesting
result is the currents along y-direction:%

\begin{align*}
J_{2y}^{\pm}  &  =C^{\pm}\{\mp(\left\vert a_{2}^{\pm}\right\vert
^{2}+\left\vert d_{2}^{\pm}\right\vert ^{2})q\ +2k_{y}\operatorname{Re}%
(a_{2}^{\pm}d_{2}^{\pm\ast}e^{2i(\theta_{\pm}+k_{r}x)})\}e^{-2k_{i}x}\\
&  =C^{\pm}\{\mp(\left\vert a_{2}^{\pm}\right\vert ^{2}+\left\vert d_{2}^{\pm
}\right\vert ^{2})\sqrt{k_{y}^{2}+k_{r}^{2}}+2\left\vert a_{2}^{\pm
}\right\vert \left\vert d_{2}^{\pm}\right\vert k_{y}\cos(2\theta_{\pm}%
+2k_{r}x+\phi_{a}^{\pm}-\phi_{d}^{\pm})\}
\end{align*}

where $\phi_{p}^{\pm}=Arg(p_{2}^{\pm})$ ($p=a,d$). Above expressions contain a
wealth of information. It is not difficult to see that for any values of
coefficients $a_{2}^{\pm}$ and $d_{2}^{\pm}$ the first term always dominates
so the currents of the two valleys are not only in opposite directions, their
directions are fixed and do not change with sign of $k_{y}$. This is unusual
and unlike the case of ordinary systems described by usual Schrodinger
equation where in case of invariance along y-direction y-component of current
$J_{y}^{S}$ always follow the sign of the $k_{y}$ and can always be written as
$J_{y}^{S}=\frac{\hbar k_{y}}{m}\rho(x)$ where $\rho(x)$ is probability
density and other factors have obvious meanings. Above result shows that, on
one hand we can use these states for valley dependant functionalities by
applying a voltage difference along the interface and on other hand, coupled
with the fact that current has to be continuous at the interface it implies
that the states of the two valleys in CR propagating towards the interface at
any positive or negative angle will carry currents in the same fixed
directions close to the interface on its \emph{both} sides. Far from the
interface in CR flux have to follow the sign of $k_{y}$. The strip-like region
around the interface where valley dependant properties show up are defined by
localization lengths of the evanescent states on both sides. This also shows
that the localized states in IR at $E$ inside the band gap modify the behavior
of those in CR that usually do not distinguish between the two valleys and
currents carried by them follow the sign of $k_{y}$. Another interesting point
is that $J_{2y}^{\pm}$ is non-zero even when $k_{y}=0$ and more important for
practical purposes is that the factors $C^{\pm}$ switch sign with the polarity
of the gates. This is clear from their expressions where all terms on right
side are positive definite except $(U-E)$ which is positive for $U>0$ and
negative otherwise. Thus we can control the directions of the currents of the
two valleys with the polarity of the gates. This broadens the possible set of
functionalities we can achieve using these states. Further, since all states
with energy inside the band gap have similar behavior, smearing of fermi
surface due to temperature or any intravalley scatterings, which are almost
unavoidable in real systems, is harmless so experimental realization of a
device where valley based operations could be performed is expected to be very
simple and easy using these localized states. One way is, similar to
ref\cite{c9} as mentioned earlier, to use just the narrow strip region of the
interface where a small voltage difference along it can be used to get a net
valley polarized current or filter a valley which can be of particles
belonging to either valley depending on the polarity of gates used to create
the band gap. Two such set ups in series can be used as a valley valve where
particles can only pass if the polarities of the gates are the same for the
two. Another way is to use the CR as the source of particles and again in this
case depending on the polarity of the gates one of the contacts on the lateral
sides will collect only particles of one valley. These contacts do not need to
be very fine on the IR, however, they must not extend in CR more than the
localized states otherwise they will also collect particles of the other
valley. And similar to the above case a combinations of such set ups can be
used for other functionalities.

For completeness, let's compute the currents $J_{2y}^{\pm}$ carried by the
localized states on IR as well as the currents along the interface
$J_{1y}^{\pm}$ in CR when the particles\ in CR travel towards the interface at
a positive energy $E$ and with y-component of wavevector $k_{y}$. The
wavefunctions in CR in this case can be written as $\Psi_{1}^{\pm}%
(x,y)=\Psi_{1}^{\pm}(x)e^{ik_{y}y}$ with $\Psi_{1}^{\pm}(x)=%
\begin{pmatrix}
1\\
-e^{\pm2i\phi}%
\end{pmatrix}
e^{+ikx}+b_{1}^{\pm}%
\begin{pmatrix}
1\\
-e^{\mp2i\phi}%
\end{pmatrix}
e^{-ikx}+c_{1}^{\pm}%
\begin{pmatrix}
1\\
h^{\pm}%
\end{pmatrix}
e^{+\kappa x}$ where $b_{1}^{\pm}$ and $c_{1}^{\pm}$ are complex coefficients,
$h^{\pm}=(\sqrt{1+\sin^{2}\phi}\mp\sin\phi)^{2}=1/h^{\mp}$, $k=\sqrt
{E-k_{y}^{2}}$ , $\kappa=\sqrt{E+k_{y}^{2}}$ and $\phi=\tan^{-1}(\frac{k_{y}%
}{k})$. Using the matching conditions $\Psi_{1}^{\pm}(x=0)=\Psi_{2}^{\pm
}(x=0)$ and $\partial_{x}\Psi_{1}^{\pm}(x)|_{x=0}=\partial_{x}\Psi_{2}^{\pm
}(x)|_{x=0}$ we determine all the coefficients $b_{1}^{\pm},c_{1}^{\pm}%
,a_{2}^{\pm}$ and $d_{2}^{\pm}$. As expected for energies inside the band gap,
$\left\vert b_{1}^{\pm}\right\vert =1$ so there is no net current along
x-direction. Further, we find that $\left\vert a_{2}^{\pm}\right\vert
=\left\vert d_{2}^{\pm}\right\vert $ $\equiv\left\vert a^{\pm}\right\vert $ so
we can write $J_{2y}^{\pm}=2\frac{\hbar\left\vert k_{y}\right\vert }{m}%
\frac{(U-E)R_{\pm}}{\Omega_{\pm}^{2}}\left\vert a^{\pm}\right\vert ^{2}%
\{\mp\sqrt{1+\left(  k_{r}/k_{y}\right)  ^{2}}+sgn(k_{y})\cos(2\theta_{\pm
}+2k_{r}x+\phi_{a}^{\pm}-\phi_{d}^{\pm})\}e^{-2k_{i}x}$. This expression
clearly shows the independence of directions of $J_{2y}^{\pm}$ from the sign
of $k_{y}$ and other features mentioned above. Currents of the two valleys in
CR $J_{1y}^{\pm}$ are determined by using $\Psi_{1}^{\pm}(x,y)$ with above
calculated coefficients $b_{1}^{\pm}$ and $c_{1}^{\pm}$. Figure(1) shows the
contour plots of the currents of the two valleys along the y-direction
$J_{y}^{\pm}$, where $J_{y}^{\pm}$ equals $J_{1y}^{\pm}$ for $x<0$ and
$J_{2y}^{\pm}$ otherwise, for $E=17meV$ and $U=\pm50meV$ as a function of
$k_{y}$ and distance from the interface. Purple (dark) and off-white (bright )
colors show currents along $-\widehat{y}$ and $+\widehat{y}$ respectively.
Figures(1a,b) show $J_{y}^{\pm}$ for $U=\pm50meV$ and $U=\mp50meV$, also
indicating that $J_{y}^{+}$ \ is the same for a given polarity of gates as
$J_{y}^{-}$ for the opposite polarity and this relation holds at all positions
$x$ and for all values of $k_{y}$. It is clear that for $U=+50meV$, close to
the interface, $J_{y}^{+}$ flows towards $-\widehat{y}$ for all possible
\emph{positive} and negative values of $k_{y}$ where as $J_{y}^{-}$ flows
towards $+\widehat{y}$ for all possible positive and \emph{negative} values of
$k_{y}$. Further, for the opposite polarity of gates, i.e., for $U=-50meV$,
currents of both valleys switch directions. From the figure, the width of the
strip-like region where this behavior is observed is approximately $15nm$ for
the parameters used so for an interface sharp enough that \ these localized
states can be used but smooth enough that intervalley scatterings can be
ignored is clearly possible. Far from the interface currents for the two
valleys retain the usual behavior. In IR they vanish when $k_{i}x>>1$ where as
in CR for $\kappa\left\vert x\right\vert >>1$ their directions follow the sign
of $k_{y}$. We can also see this fact by calculating $J_{1y}^{\pm}$ far from
the interface where the localized states have negligible effects. They are
given by $J_{1y}^{\pm}=\frac{4\hbar k_{y}}{m}\{1-\cos(2kx-2\phi-\phi_{b}^{\pm
})\}$ where $\phi_{b}^{\pm}=Arg(b_{1}^{\pm})$ which in general may be
different for the two valleys. These expressions clearly show that far from
the interface in CR currents of both valleys have the same direction
determined by the sign of $k_{y}$. In figure(2c) sum of the two currents is
plotted which shows that angular symmetric incidence of the particles of the
two valleys will result in no net charge current just like the case of common
systems, however, of course, there will be two non-zero valley currents
flowing in opposite directions.

Let's summarize above findings and draw some conclusions. We see that the
wavefunction matching conditions remain the same in case of energy dependant
effective mass in gaped bilayer graphene system which is very unusual. We
showed that the states localized near an interface between conducting and
insulating regions of BLG on the insulating side despite having plane wave
parts along the decay direction carry no currents along it. Further, we have
seen that their properties in transverse direction are strikingly different
than those of evanescent waves in ordinary systems where direction of current
follow the sign of corresponding wavevector component. Moreover, the states
belonging to the two valleys show contrasting behavior and their presence also
modifies the properties of the localized states on the conducting side of the
interface. Their valley dependent properties are insensitive to various
mechanisms that can possibly pose problems for earlier proposals to obtain
basic valley based functionalities. They are easy to exploit in realistic
situations to obtain valley polarization, filter particles of a desired valley
or for valley switching \emph{without} dealing with the propagation angles of
the particles or facing any other experimental challenges. Thus we expect them
capable of playing an important role in valleytronics based on bilayer
graphene system.

\begin{acknowledgments}
I am grateful to Emilio Artacho and Rukhshanda Naheed for helpful discussions
and Islamic Development Bank for financial support.
\end{acknowledgments}

\textbf{Figure captions:}

\textbf{Fig.1}(color online). Variation of currents $J_{y}^{\pm}$ of the two
valleys $K(+)$ and $K^{\prime}(-)$ with distance from the interface $x$ and
component of wavevector along the interface $k_{y}.$ Figure shows that for a
given polarity of gates $J_{y}^{+}$ and $J_{y}^{-}$ have fixed opposite
directions that do not depend on the sign of $k_{y}$. \textbf{(a)} $J_{y}%
^{\pm}$ for $U=\pm50meV$ \textbf{(b)} $J_{y}^{\mp}$ for $U=\pm50meV$ and
\textbf{(c)} sum of the two currents $J_{y}^{\pm}$, $J_{y}^{+}+J_{y}^{\_}$ for
$U=\pm50meV$. From the figure the width of the region that supports the valley
polarized currents along the fixed directions is roughly $15nm$.

\bigskip
\end{document}